\documentclass[12pt, a4paper]{article}

\usepackage [english]{babel}
\usepackage[dvips]{graphicx}
\usepackage{amsmath}

\date{}

\title{Gravity Probe B Experiment in 7D Space-and-Time Continuum}
\author{Yu.A. Portnov\footnote{Nuclear Safety Institute (IBRAE RAS), Russia, 115191 Moscow,
B.Tulskaya str., 52.; portnov@ibrae.ac.ru}}

\begin{document}

\maketitle

\begin{abstract}
This study deals with application of field equations in
seven-dimensional space-and-time continuum to calculate geodetic
and orbital gyroscope precession. It was demonstrated that unlike
the classic theory the assumptions made completely correspond to
the Gravity Probe B findings.
\end{abstract}

\section{Introduction}

Gravity Probe B is a space mission of the USA aimed at measuring
extremely low effects of geodetic precession of the gyroscopes on
circumterrestrial orbit which are predicted by Einstein's
relativity theory. The satellite was launched on April 20, 2004
while data acquisition process commenced in August 2004. The
satellite was for total 17 months on its orbit and completed its
mission on October 3, 2005. Reduction was completed in May 2011.
The experimental findings were published in the final report
\cite{s1}. The geodetic gyroscope precession effect coincided with
the estimated value with accuracy of 0.1\%. The effect caused by
the interaction between spin and orbital moment differs from the
estimated value for more than 19\%. This study suggests excluding
the effect caused by spinning of bodies from the design equations.

\section{Main part}

As shown in equations \cite{s6} the dynamics of both translation
and spin motion of bodies in gravity fields may be explained using
seven-dimensional space-and-time continuum which in addition to
time and three spatial coordinates comprises three coordinates
that orient a body in space and may be described as Euler angles
$x^{4}=\varphi$, $x^{5}=\psi$, $x^{6}=\theta$. For flat space the
metrics of a spherical body in an empty seven-dimensional
space-and-time continuum is given by

\begin{eqnarray}
g_{00}=1, \quad g_{\alpha\alpha}=-1, \quad g_{44}=-\frac{J}{m},
\nonumber\\ g_{45}=g_{54}=-\frac{J\cos(\theta)}{m}, \nonumber\\
g_{55}=-\frac{J}{m}, \quad g_{66}=-\frac{J}{m}, \label{e1}
\end{eqnarray}
where $J$ is the sample body inertia relative to spin axes,
precession and nutation, $m$ is the body mass, $\alpha=1,2,3$.

Equations \cite{s6} show that metrics (\ref{e1}) allows for
obtaining three classical equations of sample spherical body
motion in empty space

\begin{eqnarray}
a_{x}=0, \quad a_{y}=0, \quad a_{z}=0,  \label{e2}
\end{eqnarray}
and three gyroscope equations which cannot be derived by the
general four-dimensional relativity theory

\begin{eqnarray}
\varepsilon_{\varphi}=\frac{\omega_{\psi}\omega_{\theta}}{\sin\theta}-\frac{\cos\theta}{\sin\theta}\omega^{4}\omega_{\theta},
\nonumber\\
\varepsilon_{\psi}=\frac{\omega^{4}\omega_{\theta}}{\sin\theta}-\frac{\cos\theta}{\sin\theta}\omega_{\psi}\omega_{\theta},
\nonumber\\
\varepsilon_{\theta}=-\omega^{4}\omega_{\psi}\sin\theta.
\label{e3}
\end{eqnarray}

Gravity equations in seven-dimensional space-and-time continuum
may be written as

\begin{eqnarray}
\label{e4} R_{mn}=\ae (T_{mn}-\frac{1}{2}g_{mn}T)+\Lambda_{mn},
\end{eqnarray}
where $\Lambda_{mn}$ is supplementary tensor \cite{s6}. Variable
$\Lambda_{mn}$ must be introduced because not all components of
$R_{mn}$ with metrics (\ref{e1}) are reduced to zero in the
absence of matter $T_{mn}=0$.

Let us project the body spin vector $\omega^{4}$ on the coordinate
space

\begin{eqnarray*}
\vec{\omega}^{4}=\omega^{4}_{\alpha}dx^{\alpha}
\end{eqnarray*}
and consider translation of vector $\vec{\omega}^{4}$ along the
geodetic coordinate

\begin{eqnarray}
\label{e5}
\frac{d\omega^{4}_{\alpha}}{dt}=\Gamma^{\mu}_{\alpha\beta}u^{\beta}\omega^{4}_{\mu},
\end{eqnarray}
where $\alpha,\beta,\mu = 1, 2, 3$.

As it is shown in \cite{s2} expand field equation (\ref{e4}) into
series by velocity exponents. The, using translation equation
(\ref{e5}) we derive the gyroscope angular velocity variation
equation:

\begin{eqnarray}
\label{e6}
\frac{d\vec{\omega^{4}}}{dt}=\vec{\Omega}\times\vec{\omega^{4}},
\end{eqnarray}
where
\begin{eqnarray}
\label{e7} \vec{\Omega}=\frac{c}{2}(\vec{\nabla}\times
\vec{\xi})-\frac{3}{2}(\vec{u}\times\vec{\nabla}\phi)
\end{eqnarray}
is angular precession velocity, $\phi$ and $\xi$ are gravity
potentials. Herewith due to orthogonality of its components the
angular precession velocity may be expressed as the sum of two
orthogonal vectors

\begin{eqnarray*}\vec{\Omega}'=\vec{\Omega}'+\vec{\Omega}'',\end{eqnarray*}
where
\begin{eqnarray*}\vec{\Omega}'=-\frac{3}{2}(\vec{u}\times\vec{\nabla}\phi)\end{eqnarray*}
is angular precession velocity,

\begin{eqnarray*} \vec{\Omega}''=\frac{c}{2}(\vec{\nabla}\times \vec{\xi})
\end{eqnarray*} is angular velocity of spin-spin interaction of
the gyroscope and the Earth. Thus, the angular acceleration of the
gyroscope may be expressed as

\begin{eqnarray}
\label{e8} \varepsilon'=|\vec{\Omega}'\times\vec{\omega^{4}}|,
\end{eqnarray}
\begin{eqnarray}
\label{e9} \varepsilon''=|\vec{\Omega}''\times\vec{\omega^{4}}|.
\end{eqnarray}

Alternatively, using gyroscope equation (\ref{e3}) with the
assumption that the gyroscope angular velocity is
$\omega^{4}=const$ it is possible to derive the equations for
angular precession acceleration

\begin{eqnarray}
\label{e10}
\varepsilon_{\psi}=\omega^{4}\omega_{\theta}\sin\theta,
\end{eqnarray}
and angular nutation acceleration
\begin{eqnarray}
\label{e11}
\varepsilon_{\theta}=-\omega^{4}\omega_{\psi}\sin\theta.
\end{eqnarray}

Using the properties of spin, angular precession and angular
nutation velocity vectors \cite{s6}, and comparing equations
(\ref{e8}), (\ref{e9}), (\ref{e10}) and (\ref{e11}) we discover
that

\begin{eqnarray}
\label{e12} \omega_{\psi}=-\frac{\Omega'}{\sin\theta},
\end{eqnarray}
\begin{eqnarray}
\label{e13} \omega_{\theta}=\Omega''.
\end{eqnarray}

Henceforth we deem values $\omega_{\psi}$ and $\omega_{\theta}$
obtained in Gravity Probe B experiment to be angular velocities of
the gyroscope axis displacement.

\begin{figure}[h]
\begin{center}
\includegraphics[width=13cm]{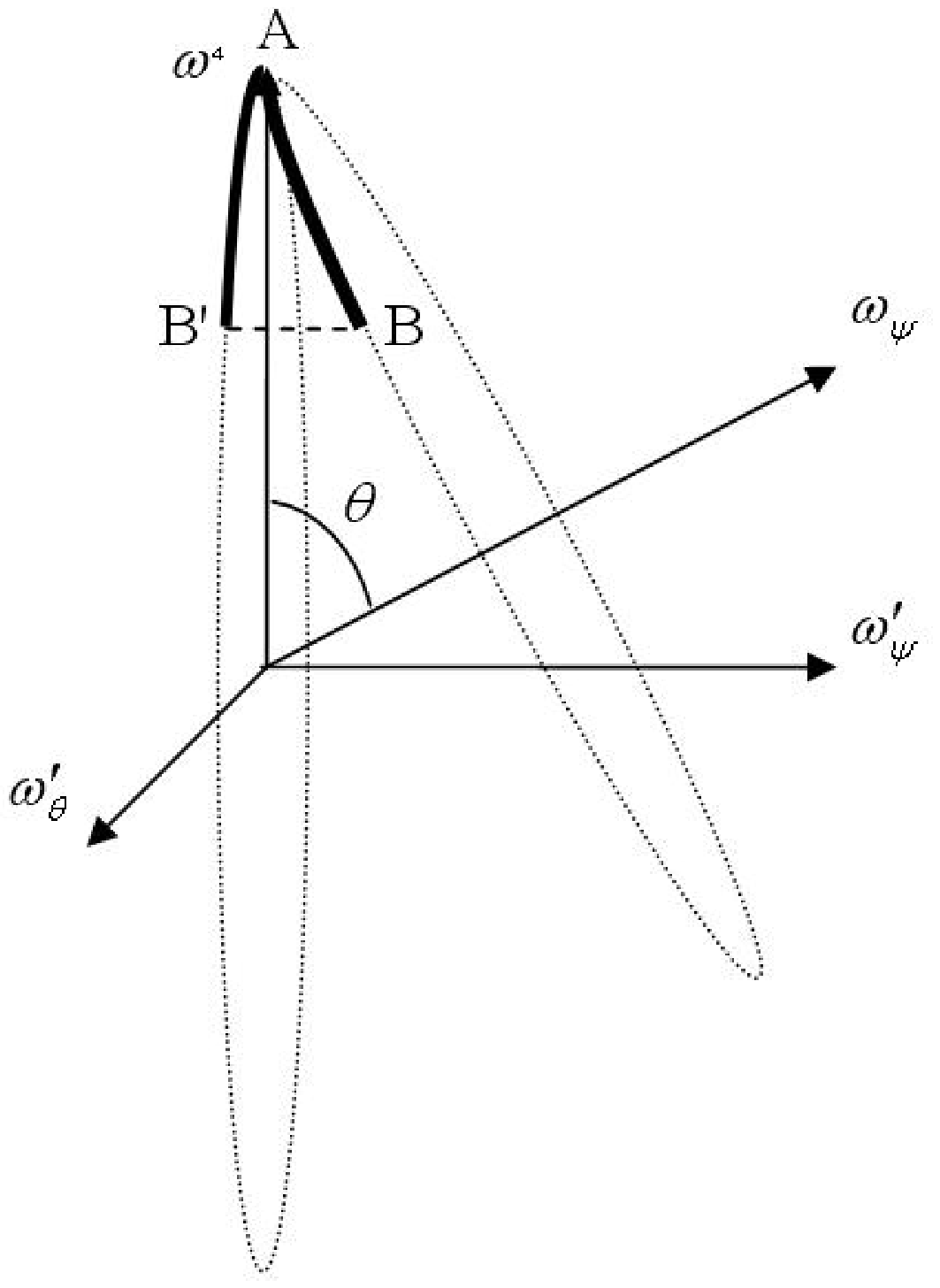}
\caption{Angular velocity vectors of spin $\omega^{4}$, precession
$\omega_{\psi}$ and virtual angular velocity vectors of precession
$\omega'_{\psi}$ and nutation $\omega'_{\theta}$} \label{fig1}
\end{center}
\end{figure}

In the performance of Gravity Probe B experiment the angular
precession velocity was expected to be orthogonal to the gyroscope
spin velocity ($\theta=\pi/2$). But should it be wrong than due to
recession it is impossible to uniquely determine by a small motion
arc of spin axis $AB$ if the arc occurs exclusively due to
nonorthogonality of precession or due to a complex motion of
orthogonal precession $AB'$ and orthogonal nutation $B'B$ (see
Fig. \ref{fig1}). That is, the angular precession velocity at may
be expended into two orthogonal components

\begin{eqnarray*}
\vec{\omega}_{\psi}=\vec{\omega'}_{\psi}+\vec{\omega'}_{\theta}
\end{eqnarray*}
of virtual precession velocity $\vec{\omega}'_{\psi}$ and virtual
nutation velocity $\vec{\omega}'_{\theta}$.

If we determine the function of virtual precession
$\omega'_{\psi}$ and virtual nutation $\omega'_{\theta}$ to actual
precession $\omega_{\psi}$ and nutation angle $\theta$ (see Fig.
\ref{fig1}) we can derive the equation

\begin{eqnarray}
\label{e14}
\omega_{\psi}'=\frac{\omega_{\psi}(\sin\theta)^{2}\sin\psi}{\sqrt{1-(2(\sin(\psi/2))^{2}(\sin\theta)^{2}-1)^{2}}},
\end{eqnarray}
\begin{eqnarray}
\label{e15}
\omega_{\theta}'=\frac{\omega_{\psi}\sin\theta\sin\psi(1-2(sin(\theta/2))^{2})}{1-\sin\theta(\sin\psi)^{2}}.
\end{eqnarray}

Substituting (\ref{e12}) into resulting relations (\ref{e14}) and
(\ref{e15}), we have
\begin{eqnarray}
\label{e16}
\omega_{\psi}'=\frac{\Omega'\sin\theta\sin\psi}{\sqrt{1-(2(\sin(\psi/2))^{2}(\sin\theta)^{2}-1)^{2}}},
\end{eqnarray}
\begin{eqnarray}
\label{e17}
\omega_{\theta}'=\frac{\Omega'\sin\psi(1-2(sin(\theta/2))^{2})}{1-\sin\theta(\sin\psi)^{2}}.
\end{eqnarray}

Therefore, we maintain that
\begin{eqnarray}
\label{e18} \Omega_{\psi}'=\omega_{\psi}',
\end{eqnarray}
\begin{eqnarray}
\label{e19} \Omega_{\theta}''=\omega_{\theta}'+\omega_{\theta}.
\end{eqnarray}
were measurable values in Gravity Probe B experiment.

\begin{figure}[h]
\begin{center}
\includegraphics[width=13cm]{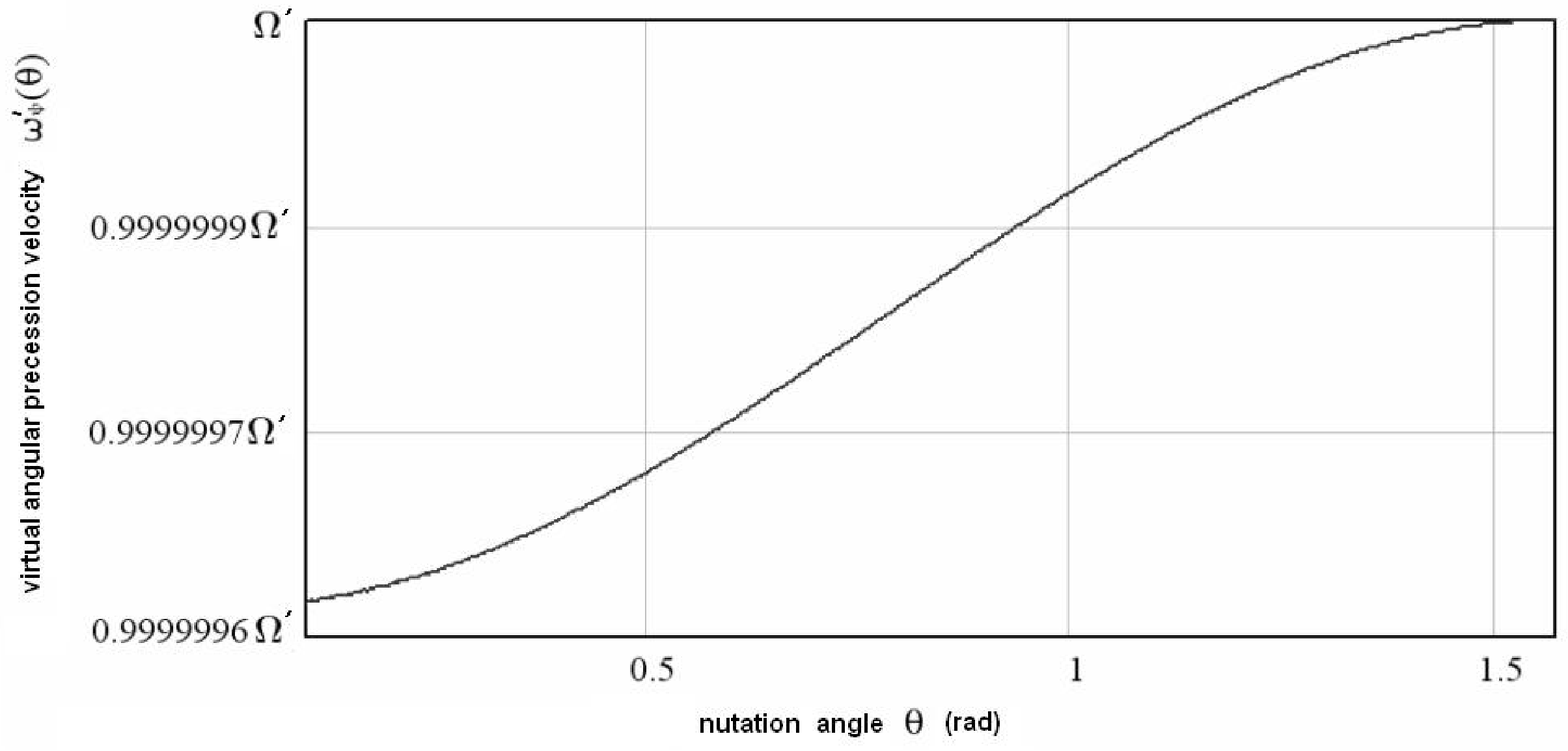}
\caption{Function of virtual angular velocity of precession
$\omega'_{\psi}$ to nutation angle $\theta$} \label{fig2}
\end{center}
\end{figure}

\begin{figure}[h]
\begin{center}
\includegraphics[width=13cm]{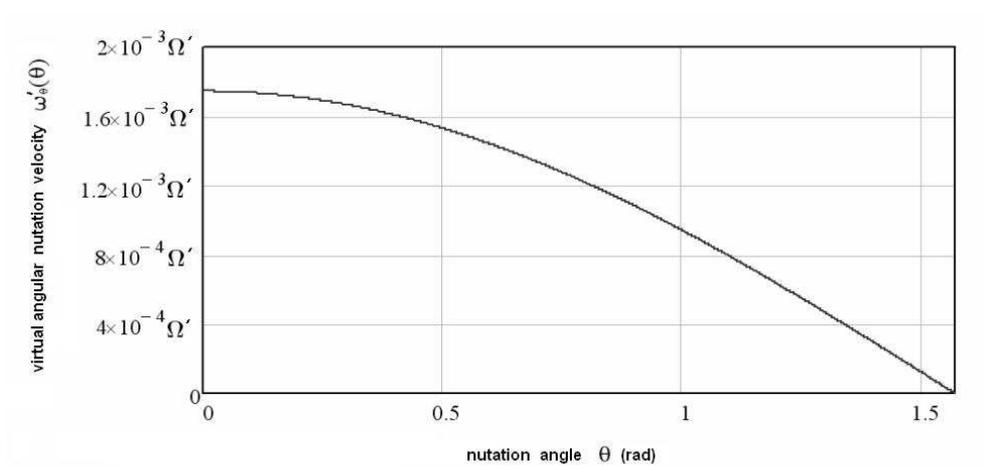}
\caption{Function of virtual angular velocity of nutation
$\omega'_{\theta}$ to nutation angle $\theta$} \label{fig3}
\end{center}
\end{figure}

As it appears from \cite{s1} the angular velocity of spin-spin
interaction differs from the actual value for about 19\%. The
functions (Fig. \ref{fig2}, Fig. \ref{fig3}) show that the
discrepancy between the theoretical and the experimental data may
be avoided if we assume that the angular nutation velocity is
contributed by misinterpretation of the angular precession
velocity, that is the difference of nutation angle $\theta$ from
$\pi/2$. In particular, at $\theta\approx 1.3965$ radian the
estimated virtual angular velocity of nutation is
$\omega'_{\theta}=3.027\cdot 10^{-4}\Omega'$ which is equal to
$\omega'_{\theta}\approx 2$ angular milliseconds per year. Thus,
if the measurable angular velocity of nutation $\Omega'_{\theta}$
is equal to the sum of a theoretical value of angular nutation
velocity $\omega_{\theta}$ (refer to (\ref{e13})) and a virtual
value of angular precession velocity $\omega'_{\theta}$ than the
estimated value shall be exactly the same as the measurable value.
Deviation of angular precession velocity $\Omega'_{\psi}$ from the
estimated value at nutation angle $\theta\approx 1.3965$ radian
will not exceed $10^{-6}$\%.

\section{Conclusion}

Finally it should be noted that considering the Gravity Probe B
experiment in terms of a seven-dimensional gravitation model it is
possible to obtain unorthogonal direction of the angular
precession velocity relatively to the gyroscope angular spin
velocity. This fact allows for interpreting the gyroscope spin
axis motion as virtual precessional and virtual nutation motions.
In view of the virtual angular nutation velocity contribution we
discover that the experimental and estimated values of angular
nutation velocity coincide at the particular angle $\theta$.

This research work has been performed in the framework of the FTsP
"Nauchnye i nauchno-pedagogicheskie kadry innovatsionnoy Rossii"
for the years 2009-2013.

\end{document}